\documentclass[aps,prl,showpacs,twocolumn,floats]{revtex4}

\usepackage{graphicx,psfig}
\usepackage{dcolumn}
\usepackage{bm}

\begin{document}

\bibliographystyle{prsty}

\title{
%
Superradiance from crystals of molecular nanomagnets \vspace{-1mm}
}

\author{
E. M. Chudnovsky$^1$ and  D. A. Garanin$^2$ }
 \affiliation{ \mbox{$^1$Department of Physics and Astronomy, Lehman College, City
University of New York,} \\ \mbox{250 Bedford Park Boulevard West,
Bronx, New York 10468-1589, U.S.A.}
\\
\mbox{$^2$Institut f\"ur Physik, Johannes-Gutenberg-Universit\"at,
 D-55099 Mainz, Germany}\\
 {\rm (Received 18 April 2002)}
}

\begin{abstract}
We show that crystals of molecular nanomagnets can exhibit giant
magnetic relaxation due to the Dicke superradiance of
electromagnetic waves.
Rigorous treatment of the superradiance induced by a field pulse
is presented.
\end{abstract}


\maketitle

High-spin molecular nanomagnets, such as spin-10 Mn$_{12}$ and
Fe$_{8}$, represent the boundary between classical and quantum
physics.
On one hand, they exhibit pronounced magnetic hysteresis, as
classical magnets do.
On the other hand, the very same magnetization curve reveals
quantum nature of spin \cite{Jonathan}.
These unique properties of molecular nanomagnets are consequence
of long-living metastable spin states \cite{Roberta} due to the
large value of spin and high energy barriers.
The lifetime of these states is believed to be dominated by
spin-phonon processes \cite{Villain96,GarChu97} and by quantum
spin tunneling \cite{Rolf,Leo,CG,Garanin91}.
The latter, for a field-sweep experiment, has been successfully
described in terms of single-molecule Landau-Zener (LZ) transitions
\cite{LZ,Zvezdin-LZ,Gunther-LZ,LL-LZ,GarChu-disl,GarSch-LZ,Wen-LZ,Sarachik-LZ}.

Recent ESR experiments
\cite{Barra-ESR,Hernandez-ESR,Barco-ESR,Sushkov-ESR,Parks-ESR}
have demonstrated noticeable resonant absorption of
electromagnetic radiation by molecular magnets.
In this Letter we show that crystals of magnetic molecules can
also be a powerful source of coherent electromagnetic radiation.
At low fields, the magnetic relaxation of molecular nanomagnets
may be affected by the distribution of energy levels due to
dipolar fields \cite{Stamp}, nuclear spins \cite{Stamp,GarChuSch},
and crystal defects \cite{GarChu-disl}.
Here we will study the case when the resonant tunneling of the
spin of individual molecules is dominated by the large external magnetic field
(or by a large transverse anisitropy).
We will be interested in relaxation between the tunnel-splitted ground states
of magnetic molecules on the two sides of the barrier which has two specific features.
Firstly, the change of the projection of the spin on the symmetry axis, $S_z$,
is of order $S\gg 1$ and the coupling with the light is increased by a factor of $S$ in
 comparison with the usual processes in which $S_z$ changes by one.
Secondly, phonon relaxation processes between these states appear only at high orders
of the perturbation theory and are strongly suppressed \cite{chumar02} hence
electromagnetic relaxation will prevail.

One might think that the individual relaxation of
molecules is always a good approximation.
Note, however, that in molecular nanomagnets the wavelength
$\lambda$ of the electromagnetic radiation due to transitions
between spin levels is typically greater than the dimensions of
the crystal $L$.
Consequently, the molecules can coherently interact through the
electromagnetic radiation that they are emitting, thus greatly
enhancing its intensity \cite{Dicke}.
\begin{figure}[t]
\unitlength1cm
\begin{picture}(11,6)
\centerline{\psfig{file=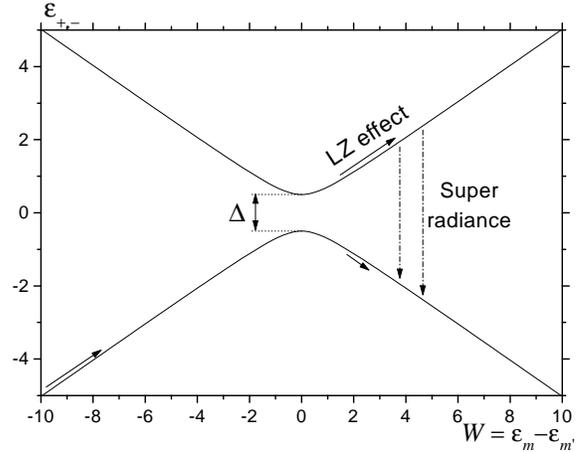,angle=-90,width=9cm}}
\end{picture}
\caption{ \label{fig_spl}
A pair of tunnel-splitted levels vs. energy bias $W$.
Coherent light is emitted after crossing the resonance via superradiance.
}
\end{figure}
\begin{figure}[t]
\unitlength1cm
\begin{picture}(11,6)
\centerline{\psfig{file=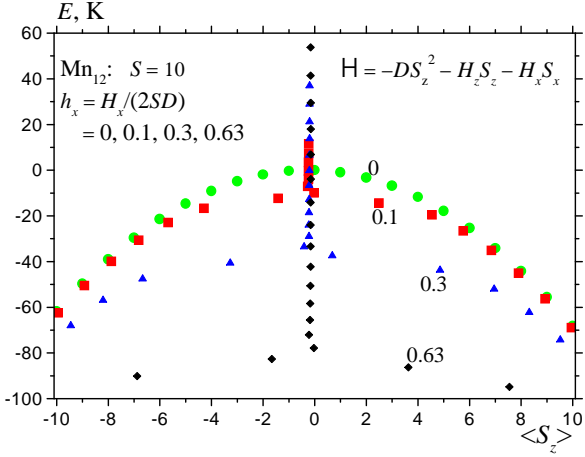,angle=-90,width=9cm}}
\end{picture}
\caption{ \label{fig_levels}
Energy levels of the Mn$_{12}$ Ac molecule for different transverse fields $H_x$ off resonance ($g\mu_B H_z=D/2$), labeled by
$\langle S_z\rangle$.
States with $\langle S_z\rangle\approx 0$ are not localized in one of the potential wells.
$h_z\simeq 0.63$ ($H_x\simeq 6\,$T) generates the ground-state splitting $\Delta/k_B\simeq 0.15\,$K.
 }
\end{figure}

Consider a crystal of molecular magnets (e.g., Mn$_{12}$ or
Fe$_{8}$) of spin $S$, with a large
energy barrier between spin-up and spin-down states, near a
resonant value of the magnetic field $H_z$.
Since the tunnel splitting ${\Delta}$ is small compared to
other energy distances, the pair of splitted energy eigenvalues can be approximated
by (see  Fig.\ \ref{fig_spl}).
\begin{equation} \label{epspmDef}
\varepsilon _{\pm }=\frac{1}{2}\left( \varepsilon _{2}+\varepsilon
_{1}\pm \sqrt{W^{2}+\Delta ^{2}}\right),
\end{equation}
where $W\equiv \varepsilon_{1}-\varepsilon _{2}$, and the
eigenstates are given by
\begin{equation}\label{psipm}
{\psi}_{\pm}=\frac{1}{\sqrt{2}}\left(\pm C_\pm \psi_1 +
C_{\mp}\psi_2\right)
\end{equation}
with $C_{\pm }\equiv \sqrt{1\pm W/\sqrt{\Delta^2+ W^2}}$
\cite{GarSch-LZ}. Here $\psi_1$ and $\psi_2$ are eigenstates in
the left and right wells with tunneling neglected.
If the transverse field $H_x$ is small, the states $\psi_1$ and $\psi_2$
 are just $|S\rangle$ and $-|S\rangle$.
In the strong field the ground states of the system in the wells are described by the tilted spin
 and its projection $\langle S_z\rangle_0$.
In this case
\begin{equation}\label{psipm}
 W=2\langle S_z\rangle_0 g\mu_B H_z.
\end{equation}
For $S\gg 1$ the classical approximation for $\langle S_z\rangle_0$ works well.
In particular, for Mn$_{12}$ (${\cal H} = -D S_z^2 -g\mu_B H_x S_x$)  one has
\begin{equation}\label{mClass}
\langle S_z\rangle_0 \approx S \sqrt{1-h_x^2}, \qquad h_x\equiv g\mu_B H_x /(2SD).
\end{equation}
To create an appreciable tunnel splitting, one has to apply a strong transverse field
(e.g., $H_x\simeq 6\,$T for  $\Delta/k_B=0.15\,$K for Mn$_{12}$).
For such values of $H_x$ potential wells become shallow (see Fig.\ \ref{fig_levels})
 and most of the levels are no longer localized in the wells; still
  $\langle S_z\rangle_0 \approx 0.777S$ is large.

We assume that the crystal is in thermal equilibrium with the
electromagnetic radiation and at first compute the rate of
transition between ${\psi}_{-}$ and ${\psi}_{+}$ for a single
molecule.
Interaction with the quantized radiation field ${\bf H}_{\rm pht}$ is
${\hat{V}}=-g{\mu}_{B}{\bf H}_{\rm pht}{\bf S}$
\cite{Landau}.
(Notice that only the $z$-component of ${\bf H}_{\rm pht}$ couples
to the tunneling system.)
The decay rate is given by the Fermi golden rule
\begin{equation}\label{FermiGoldenRule}
{\Gamma}_{1}=\frac{4}{3\hbar}|\langle{\psi}_{+}|S_{z}|{\psi}_{-}\rangle|^{2}
(g \mu_B)^2 k^3 \coth\left(\frac{{\hbar}{\omega}}{2k_B T}\right),
\end{equation}
where $k=n\omega/c$ is the wave vector of a photon,
$n$ is the refraction index of the crystal, and
\begin{equation}
\omega=\sqrt{\Delta^2+W^2}/\hbar
\end{equation}
is the photon frequency.
The matrix element squared is
\begin{equation}\label{MatrixElement}
|\langle{\psi}_{+}|S_{z}|{\psi}_{-}\rangle|^{2}=\langle S_z\rangle_0^2\frac
{\Delta^2}{\Delta^2+W^2},
\end{equation}
making ${\Gamma}_1$ proportional to $\omega\Delta^2$.
When $k_B T \lesssim \Delta$ and $\Delta/\hbar$ is as high as
$2{\times}10^{10}\,$s$^{-1}$ (i.e., $\Delta/k_B\simeq 0.15\,$K),
Eq.\ (\ref{FermiGoldenRule}) at $W \leq \Delta$ gives $\Gamma_1
\sim 7.4\times 10^{-12}\,$s$^{-1}$, which is negligible compared with typical
experimental transition rates.
The situation changes drastically, however, when one studies
interaction of a macroscopic number $N$ of magnetic molecules with
the electromagnetic radiation.

When the wave length of emitted photons exceeds the linear size of
the sample, $\lambda \gg L$, then the phase of these photons is
the same throughout the sample and the process of radiation
becomes coherent.
Near the tunnelling resonance the system can be described by the
effective Hamiltonian
\begin{equation}\label{PseudoHam}
{\cal H_{\rm eff}} = -g\mu_B {\bf H}_{\rm eff}\cdot{\bf R} + \hat
V_{\rm eff} ,\qquad {\bf R}\equiv \sum_{i=1}^N \bm{\tau}_i
\end{equation}
where $\bm{\tau}_i$ is a fictitious spin 1/2 describing the
resonant states of the $i$-th molecule,
\begin{equation}\label{HeffDef}
g\mu_B {\bf H}_{\rm eff} = \Delta {\bf e}_x+W{\bf e}_z,
\end{equation}
$\hat V_{\rm eff}=2\langle S_z\rangle_0 g\mu_B R_z H_{{\rm pht},z}$ is the spin-photon interaction.
It can be checked that ${\cal H_{\rm eff}}$ conserves quantum
number ${\bf R}^{2}=R(R+1)$.
We are interested here in the states with the maximal value of
$R$, $R=R_{\rm max}=N/2$. Such states are produced when we begin
with all magnetic molecules in the left well in Fig.1, that is,
magnetized down, and then sweep the applied field across the
resonance.
The dynamics of the system is that of a large spin ${\bf R}$
coupled to photons, and it is convenient to label its states by
the quantum number $M$ which is the projection of ${\bf R}$ on ${\bf
H}_{\rm eff}$ of Eq.\ (\ref{HeffDef}).
The matrix elements of the Hamiltonian Eq.\ (\ref{PseudoHam}) for
transitions between $|R,M\rangle$ and $|R,M \pm 1\rangle$ contain
factors $\sqrt{(R \pm M)(R \mp M+1)}$ that are additional to those
for a single molecule.
The latter makes the spontaneous radiation power $I$ to scale as
$(R+M)(R-M+1)$.
If $|M|\ll R$ and $R\sim N$, then $I\propto N^2$, which is $N$
times the incoherent radiation from a system of $N$ molecules.
This is the superradiance discovered by Dicke \cite{Dicke}.

At finite temperatures, the time evolution of a system of
tunneling molecules in equilibrium with the radiation field is
described by the density-matrix equation (DME) for a spin ${\bf
R}$.
For $R\gg 1$ and low temperatures, $R g\mu_B H_{\rm eff}/(k_B T)=R
\hbar\omega/(k_B T)\gg 1$ (but not necessarily $\hbar\omega/(k_B
T)\gg 1$, see Ref.\ \cite{gar91llb}), this DME goes over to the
classical Landau-Lifshitz equation,
\begin{equation}\label{LL}
\dot{\bf n}=\gamma [{\bf n\times H}_{\rm eff}] - \alpha\gamma[{\bf
n}{\times}[{\bf n}{\times}{\bf H}_{\rm eff}]], \qquad {\bf
n}\equiv {\bf R}/R,
\end{equation}
where $\gamma = g\mu_B/\hbar$ is the gyromagnetic ratio,
\begin{equation}\label{alphaDef}
\alpha= \frac {R\Gamma_1 }{\gamma H_{\rm eff}} \tanh\left(\frac{g
\mu_B H_{\rm eff}}{2k_B T}\right)
\end{equation}
is dimensionless damping coefficient, $\Gamma_1$ is given by Eq.\
(\ref{FermiGoldenRule}) and ${\bf H}_{\rm eff}$ is defined by Eq.\
(\ref{HeffDef}).

The first term in Eq.\ (\ref{LL}) gives dissipationless
Landau-Zener transitions when the field is swept through the
resonance such that $W=W(t)$ satisfies $W(\pm\infty)=\pm\infty$,
 and the initial condition is
${\bf n}(-\infty)=-{\bf e}_z$.
Indeed, the Schr\"odinger equation for a two-level system is
equivalent to the equation for a precessing spin.
The probability $P(t)$ for a molecule to stay in the initial state
is given by
\begin{equation}\label{Pvianz}
P(t)=[1-n_z(t)]/2.
\end{equation}
For $W(t)=vt$, one obtains the LZ result \cite{LZ,Zvezdin-LZ}
\begin{equation}\label{LZ}
P(\infty)\equiv P=\exp[-\pi \Delta^2/(2\hbar
v)].
\end{equation}

The second term in Eq.\ (\ref{LL}) describes Dicke superradiance
since $\alpha$ contains the large factor $R=N/2$.
Eqs.\ (\ref{alphaDef}), (\ref{FermiGoldenRule}),
(\ref{MatrixElement}) and relation $g \mu_B H_{\rm
eff}=\sqrt{\Delta^2+W^2}$ yield
\begin{equation}\label{alphaRes}
\alpha=\frac 16 N\langle S_z\rangle_0^2 g^2 n^3
\left(\frac{e^2}{\hbar c}\right) \left(\frac{\Delta}{m_e
c^2}\right)^2,
\end{equation}
where $e$ is electron charge, $m_e$ is electron mass and $c$ is
the speed of light.
Note that $\alpha$ is independent of the energy bias $W$.
The corresponding longitudinal decay rate is
\begin{equation}\label{GammaDicke}
\Gamma = 2\alpha \gamma H_{\rm eff}=2\alpha \omega =2R
\Gamma_1^{(0)}=N\Gamma_1^{(0)},
\end{equation}
where $\Gamma_1^{(0)}$ is given by Eq.\ (\ref{FermiGoldenRule})
without the $\coth$ factor.
Thus superradiance boosts the one-molecule relaxation rate by a macroscopically large factor $N$.

Let us consider at first the solution of Eq.\ (\ref{LL}) for ${\bf H}_{\rm eff}={\rm const}$.
In this case it is convenient to describe
the motion of the Dicke pseudospin ${\bf n}$ in the coordinate system with the $z'$ axis along ${\bf H}_{\rm eff}$.
The physically relevant $z$ component of ${\bf n}$ is given by
\begin{equation}\label{nz}
 n_z=n_{z'}\cos\psi  - n_{x'}\sin\psi, \quad
\cos\psi=\frac W{\sqrt{\Delta^2+W^2}},
\end{equation}
whereas the solution for $n_{z'}$ and  $n_{x'}$reads
\begin{eqnarray}\label{SolLL}
&&
 n_{z'}(t)=\frac{\sinh(\tau) + n_{z'}(0)\cosh(\tau)}
{\cosh(\tau) + n_{z'}(0)\sinh(\tau)}, \quad \tau\equiv
\alpha\gamma H_{\rm eff}t, \nonumber\\
&&
 n_{x'}(t)=\sqrt{1-n_{z'}^2(t)}\sin(\gamma H_{\rm eff}t+\phi_0).
\end{eqnarray}
The full relaxation from $n_{z'}=-1$ to $n_{z'}=1$ is given by
$n_{z'}(t)=\tanh(\tau)$ which can be used to obtain the value of
$\alpha$ from an independent macroscopic argument.
To this end, we set $R=N/2$ and equate the change of Zeeman energy
$Ng\mu_B H_{\rm eff}$ [see  Eq.\ (\ref{PseudoHam})] to the energy
dissipated due to the magnetic-dipole radiation
$\int_{-\infty}^\infty I(t) dt$,
\begin{equation}\label{MDP}
I=[2/(3c^3)]\ddot m_z^2,
\end{equation}
where $m_z=N\langle S_z\rangle_0 g\mu_B n_z$.
For $\alpha\ll 1$ one obtains
\begin{equation}
\ddot n_z  \cong  -\ddot n_{x'}\sin\psi
           \cong
(\gamma H_{\rm eff})^2 n_{x'} \sin\psi.
\end{equation}
With Eq.\ (\ref{SolLL}) this yields
\begin{equation}\label{ddotnz2}
\ddot n_z^2\Rightarrow (\gamma H_{\rm eff})^4\sin^2\psi
[1-n_{z'}^2(t)]/2,
\end{equation}
and integration of Eq.\ (\ref{MDP}) using $n_{z'}(t)=\tanh(\tau)$ leads exactly to
Eq.\ (\ref{alphaRes}).
For $\alpha\gtrsim 1$ this method shows a breakdown of Eq.\ (\ref{LL})
because of the impossibility to treat the spin-photon interaction as a perturbation.

\begin{figure}[t]
\unitlength1cm
\begin{picture}(11,6)
\centerline{\psfig{file=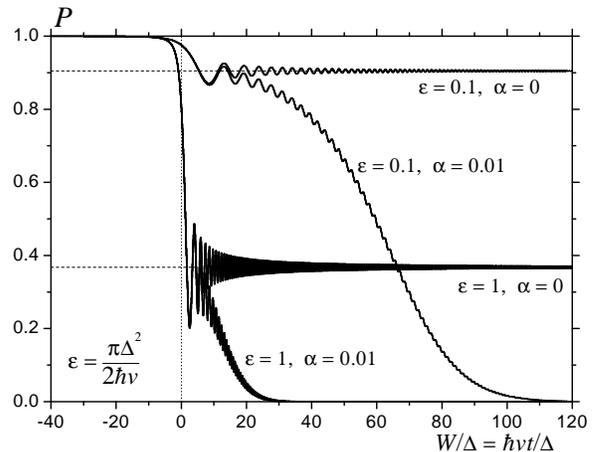,angle=-90,width=9cm}}
\end{picture}
\caption{ \label{fig_POne}
Time dependence of the probability to occupy the initial state $m$ during
the linear energy sweep with different sweep rates and damping coefficients
 $\alpha$ obtained by numerical solution of Eq.\ (\protect\ref{LL}).
}
\end{figure}

In the Landau-Zener experimental setup, the evolution of the system
described by Eq.\ (\ref{LL}) proceeds for $\alpha\ll 1$ in two stages, see Fig.\ \ref{fig_POne}.
The first stage is the LZ process that provides the
fraction of molecules $P$ given by Eq.\ (\ref{LZ}) in
the excited states at $t>0$ (the upper energy branch in Fig.\ \ref{fig_spl}).
In the second stage these excited states decay due to the
superradiance onto the lower branch in Fig.\ \ref{fig_spl}.
According to Eq.\ (\ref{LZ}), at a small sweep rate
almost all molecules follow the lower energy branch, so that the
corresponding radiation power is small.
To produce a powerful coherent electromagnetic radiation the
resonance should be crossed at a rate $v \gtrsim \Delta^2/\hbar$.
To obtain the radiation at a fixed frequency, a short longitudinal field pulse is needed to drive a crystal of
molecular magnets into an excited state with
$n_{z'}(0)=-\cos\psi$ and $n_{x'}(0)=\sin\psi$.
The time evolution of this state is described by Eqs.\
(\ref{SolLL}) and (\ref{nz}), whereas the electromagnetic
radiation is given by Eq.\ (\ref{MDP}).

Let us now analyze limitations on the superradiance relaxation rate of Eq.\ (\ref{GammaDicke}).
Firstly, the linear size of the sample $L$ should not exceed the wave length of the emitted photons $\lambda=2\pi/k$,
otherwise photons emitted by different molecules will be out of phase
 and coherence will be destroyed.
A conservative estimate is
\begin{equation}\label{kL}
kL\lesssim 1, \qquad k= n\omega/c = n\sqrt{\Delta^2+W^2}/(c\hbar)
\end{equation}
 which with $N=L^3/v_0$ yields
\begin{equation}\label{GammaMax}
\Gamma\lesssim \frac{E_d}\hbar  \frac {\Delta^2}{\Delta^2+W^2},
\qquad E_d=\frac{(\langle S_z\rangle_0 g\mu_B)^2}{v_0},
\end{equation}
where $E_d$ is the dipole-dipole energy and $v_0$ is the unit-cell volume.
The maximal rate $\Gamma_{\rm max}\sim E_d/\hbar$ and thus the maximal intensity of the electromagnetic radiation
can be achieved at the resonance, $W=0$.

Apart of that, our theory makes use of the weak coupling $\alpha\lesssim 1$
[see comment below Eq.\ (\ref{ddotnz2})].
To work out its implications, it is convenient to rewrite Eq.\ (\ref{alphaRes}) as
\begin{equation}\label{alphaAnother}
\alpha=(2/3) (k_0 L)^3,\qquad k_0\equiv n(\Delta^2 E_d)^{1/3}/(c\hbar).
\end{equation}
Thus $\alpha\lesssim 1$ converts into $k_0L\lesssim 1$.
The weak-coupling condition is related to the requirement that
the dipole-dipole interaction is sufficiently weak  to prevent the
macroscopic system from dipolar ordering at low temperature (see, e.g., \ \cite{marchuaha01}).
In the thermodynamic limit, if the condition $E_d\lesssim \Delta$ is fulfilled,
then in the ground state the pseudospins in   Eq.\ (\ref{PseudoHam}) point in the $x$ direction
instead of ordering along the $z$ axis, as favored by the DDI.
One can see that at the border of the macroscopic region, $kL\simeq 1$, Eq.\ (\ref{alphaAnother})
yields exactly $\alpha\simeq 1$ for $E_d\simeq \Delta$ and $W=0$.
For $E_d\lesssim \Delta$ the weak-coupling condition $k_0L\lesssim 1$ is less restrictive
than the coherence requirement $kL\lesssim 1$ that leads to the maximal rate of Eq.\ (\ref{GammaMax}).
If $\Delta\lesssim E_d$, then, at resonance, the condition $k_0L\lesssim 1$
is more restrictive than $kL\lesssim 1$,
and the estimation $\Gamma_{\rm max}\sim E_d/\hbar$ becomes invalid.
Note, however, that for small tunnel splittings $\Delta$, both $k$ (at resonance) and $k_0$ are
small and thus both conditions are fulfilled for realistic $L$.

For Mn$_{12}$ in the transverse field $H_x=6\,$T, so that $\langle S_z\rangle_0=0.777S$ and
$\Delta/\hbar=2{\times}10^{10}\,$s$^{-1}$  ($\Delta/k_B\simeq
0.15\,{\rm K}>E_d/k_B\simeq 0.04\,$K), Eq.\ (\ref{alphaRes}) with
 $g=2$, and $n=1$ gives $\alpha \simeq N\times 1.9 \times 10^{-22}$.
Taking $N=L^3/v_0$ with the linear size of the crystal $L=0.5\,$cm
and the unit-cell volume $v_0=3.7\times 10^{-21}\,$cm$^3$ for
Mn$_{12}$, one obtains $N=3.4\times 10^{19}$ and $\alpha\simeq
0.0063$.
Then Eq.\ (\ref{GammaDicke})  gives $\Gamma \simeq 2.5\times
10^8\,$s$^{-1}$.
For a single crystal of Mn$_{12}$, the maximal possible
superradiance rate given by Eq.\ (\ref{GammaMax}) is $\Gamma_{\max}\simeq 7\times
10^9\,$s$^{-1}$.
For $\Delta/k_B\simeq
0.15\,$K this rate is achieved at $L \simeq 1/k \simeq 1.5\,$cm which is rather large.
In this case $\alpha\simeq 0.17$,
and the weak-coupling approximation should  still be applicable.
For the crystal size $L\simeq 0.5\,$cm, the maximal rate corresponds to the
emission of photons of frequency $f=\omega/(2\pi)\simeq 10\,$GHz, i.e.,
$\Delta/k_B\simeq 0.46\,$K.
Similar estimations could be done for Fe$_8$ which is a better candidate for
observing superradiance since in Fe$_8$ the hyperfine
interactions that may cause decoherence are weak \cite{Wen-LZ}.
Taking for Fe$_8$ $v_0=2.0 \times 10^{-21}\,$cm$^3$,
one obtains $\Gamma_{\max}{\sim}1.3{\times}10^{10}\,$s$^{-1}$.
We have to stress that superradiance can be observed for much smaller
 tunnel splittings $\Delta$ than we used in the estimations above,
 with accordingly lower relaxation rates.
 The only lower limitations on $\Delta$ are that (i) it has a
narrow distribution throughout the crystal and (ii) $\Gamma$
of Eq.\ (\ref{GammaDicke}) exceeds the rates of various incoherent decay
processes.

In conclusion, we have demonstrated that crystals of
molecular nanomagnets can exhibit superradiance in the
broad frequency range.
Our theory is based upon
fundamental electrodynamics of a large spin and is
insensitive to the details of the crystal structure
as long as the spectrum of molecular spin levels consists
of narrow lines determined by the crystal field and/or
external magnetic field.
 Experimentalists should, therefore,
focus on large single crystals of weakly interacting molecular
magnets with weak hyperfine interactions.

Our thinking on this problem has been provoked by
 Javier Tejada.
This work has been
supported by the NSF Grant No. 9978882.
%



\end{document}